\documentclass[nofootinbib,notitlepage,aps,twocolumn, letterpaper,superscriptaddress]{revtex4-1}

\usepackage{graphicx}
\usepackage{morefloats}
\usepackage{color}
\usepackage{amssymb}
\usepackage{float}
\usepackage{bm}
\usepackage{amsmath}
\usepackage{epstopdf}

\usepackage[table]{xcolor}

\begin{document}

\newcommand{\ham}{\mathcal{H}} 
\newcommand{\Q}{\mathcal{Q}} 

\title{Demonstration of nanosecond operation in stochastic magnetic tunnel junctions }

\author{Christopher Safranski}
\affiliation{IBM T. J. Watson Research Center, Yorktown Heights, New York 10598, USA}
\author{Jan Kaiser}
\affiliation{IBM T. J. Watson Research Center, Yorktown Heights, New York 10598, USA}
\affiliation{Current address: School of Electrical and Computer Engineering, Purdue University, IN 47907, USA}
\author{Philip Trouilloud}
\author{Pouya Hashemi}
\author{Guohan Hu}
\author{Jonathan Z Sun}
\affiliation{IBM T. J. Watson Research Center, Yorktown Heights, New York 10598, USA}


\flushbottom
\maketitle

\textbf{Magnetic tunnel junctions operating in the superparamagnetic regime are promising devices in the field of probabilistic computing, which is suitable for applications like high-dimensional optimization or sampling problems.  Further, random number generation is of interest in the field of  cryptography. For such applications, a device's uncorrelated fluctuation time-scale can determine the effective system speed. It has been theoretically proposed that a magnetic tunnel junction designed to have only easy-plane anisotropy provides fluctuation rates determined by its easy-plane anisotropy field, and can perform on nanosecond or faster time-scale as measured by its magnetoresistance's autocorrelation in time. Here we provide experimental evidence of nanosecond scale fluctuations in a circular shaped easy-plane magnetic tunnel junction, consistent with finite-temperature coupled macrospin simulation results and prior theoretical expectations. We further assess the degree of stochasticity of such signal.   }

\begin{figure*}[tbp]
\includegraphics[width=\textwidth]{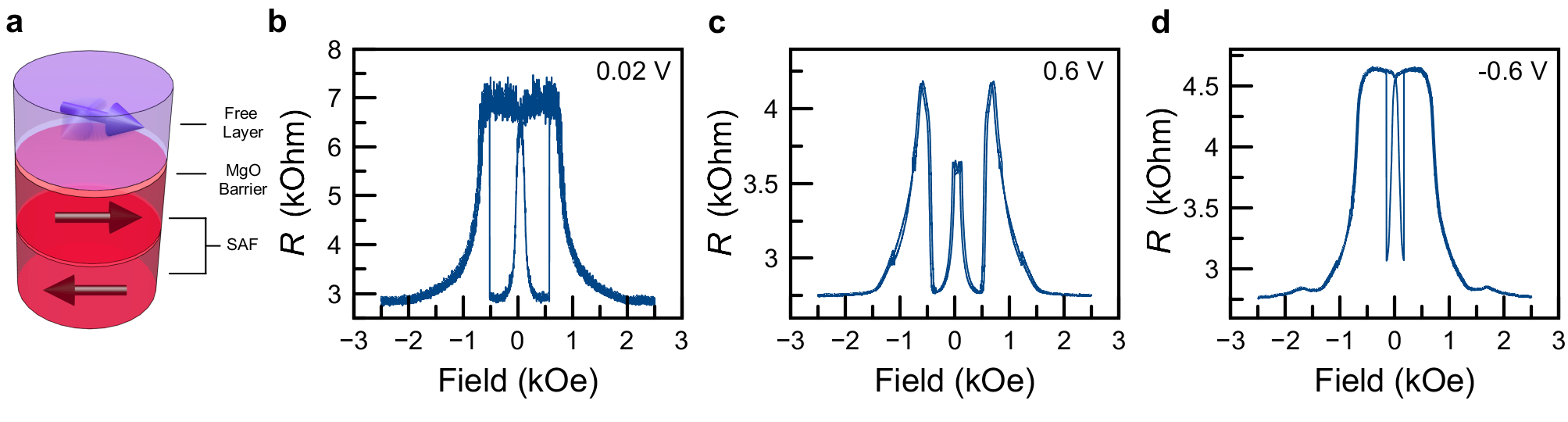}
\caption{ 
(\textbf{a}) Schematic of the magnetic tunnel junction.  Resistance vs applied field at (\textbf{b}) 0.02\,V, (\textbf{c}) 0.6\,V and (\textbf{d}) -0.6\,V. 
}
\label{fig1}
\end{figure*} 

In recent years, probabilistic computing based architectures, such as Ising and Boltzmann Machines \cite{2017150,2017053,2017084,2018112,2018113,2019132,2019133,2019136,yamamoto_73_2020,aramon_physics-inspired_2019,goto_combinatorial_2019,sutton_autonomous_2020,hamilton_accelerating_2020} have seen a revival. These offer improved efficiency compared to the traditional deterministic von Neumann approach to solving computational hard problems such as the traveling salesman problem and factorization.\cite{2018115,borders_integer_2019} These computation approaches require a large number of independent sources of stochastic signal since they are often based on Markov Chain Monte Carlo techniques like Gibbs sampling.\cite{geman_stochastic_1984} Thus an economical, low-power, high-density hardware true-random noise source is of interest. Magnetic tunnel junctions (MTJ) have already been integrated into CMOS technologies on scale for memory applications\cite{2017019,2020085}, where they are engineered to have stable magnetic states. By changing the materials and geometric design, MTJs can also become naturally fluctuating.   Determining the limiting speed of such fluctuation is important, since it relates to the speed of computation in a probabilistic computing scheme\cite{sutton_autonomous_2020}. 

Stochastic signal generation  has been reported\cite{2017150,borders_integer_2019} in  super-paramagnetic MTJs with  low uniaxial anisotropy energy barriers, operating  in the relatively slow, millisecond time regime. For macrospin in the limit $E_{b}/k_B T\geq1$, the superparamagnetic fluctuation rate follows an Arrhenius-like relation\cite{2003004,2011002B,2013104}, with the fluctuation rate $1/\tau_F \propto\left( 1/\tau_0\right) \exp\left( - E_b/k_B T\right)$. An exponential increase in fluctuation rate is expected upon lowering of the energy barrier height $E_{b}/k_B T $, with a frequency scale set approximately by the attempt frequency of $1/\tau_{o}\propto\alpha \gamma H_k$, where $\alpha$ is the macrospin's Landau-Lifshitz damping coefficient, $H_k$ the anisotropy field, and $\gamma \sim 2 \mu_B/\hbar$ the magnitude of the gyro-magnetic ratio. Since both energy barrier height $E_b=m H_k/2$ (with $m$ being the total moment of the macrospin) and attempt frequency decrease with decreasing $H_k$, there is a limit in fluctuation rate. The low anisotropy limit $E_{b}/k_B T\ll 1$ distributes magnetic states more isotropically, with a fluctuation rate of the order $\alpha \gamma k_B T/m$\cite{2003004,2020088}. For common materials and device parameters, this limiting speed is $\sim$ 10 to 100\,MHz. 
 
A potential way to increase the fluctuation speed, proposed recently\cite{2020088,2019122}, is to utilize easy-plane anisotropy that can allow super-paramagnetic fluctuation confined in-plane, while preserving some high-speed fluctuation dynamics.  In this configuration, the energy barrier is primarily set by the shape of the magnetic tunnel junction. A low energy barrier can  be achieved by constructing a circular in-plane junction. The attempt frequency for this configuration is then related to the free layer's easy-plane anisotropy field, which can be significantly higher than easy-axis anisotropy field $H_k$, allowing for a faster fluctuation. Nanosecond scale dwell times have been reported in earlier  in-plane structures with uniaxial anisotropies\cite{Pufall2004},  however,  the autocorrelation time of such structures has not been thoroughly investigated.  


In this letter, we experimentally explore an in-plane MTJ's fluctuation behavior, and measure its fluctuation speed and stochasticity.   We fabricate  MTJs  with $\sim$ 60\,nm circular diameters  based on CoFeB (Methods). The resulting device structure is represented schematically in Fig.\ref{fig1}a.   Here the top CoFeB above MgO is the free layer (FL). The reference layer (RL) below MgO consists of two CoFeB ferromagnets antiferromagnetically coupled through Ru, forming a so-called synthetically antiferromagnetic (SAF) structure for dipole field compensation. 

Figure\,\ref{fig1}b shows the quasi-static ($\ll 10$  kHz) magnetoresistance of the MTJ as a function of applied magnetic field measured at +20mV of DC bias across the MTJ. We define +bias as the direction where tunnel-current related spin-torque favors the parallel (P) alignment of the FL with respect to the RL. The resistance thus measured is the low-pass time-averaged mean value. At zero applied field, the MTJ is in the high resistance anti-parallel (AP) state, due to incomplete compensation of the dipole field from the SAF RL. At higher field, both FL and RL are nearly saturated along the applied field direction, resulting in a lower resistance near the P state. When the same measurement is performed at +0.6\,V in Fig.\,\ref{fig1}c, we observe the the magnetoresistance behavior is changed. This polarity of bias voltage results in a spin torque that encourages parallel orientation of two layers. Interestingly, we observe that the resistance near zero field is not the AP or P state resistance, but rather an in-between value. Measurement at negative voltages  in Fig.\,\ref{fig1}d show that at low field the junction tends to stay in the AP state, consistent with the sign change of spin torque. 

\begin{figure*}[tbp]
\includegraphics[width=\textwidth]{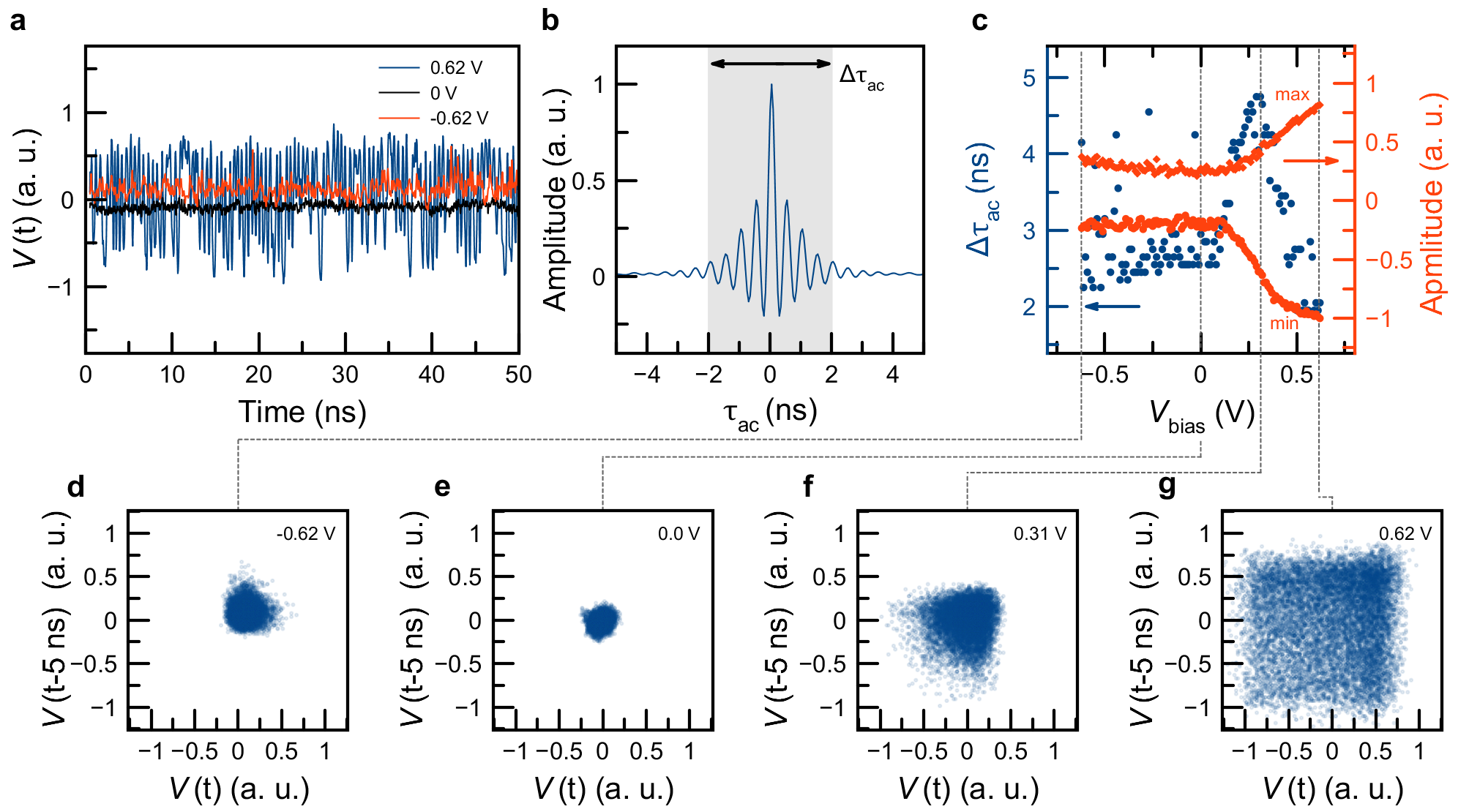}
\caption{ 
(\textbf{a}) Time traces of the MTJ signal with no bias voltage and high bias. (\textbf{b}) The autocorrelation function of the measured time traces. (\textbf{c}) The linewidth of the autocorrelation as a function of bias voltage (blue) and the max and minimum time domain signal amplitudes (red). (\textbf{d}-\textbf{g}) The time domain signal plotted against the same signal delayed 5\,ns for the voltages shown by the dotted lines in (\textbf{c}). The scales for signal voltage in (\textbf{a}),(\textbf{c}) and (\textbf{d}-\textbf{g}) are normalized against the highest amplitude obtained at +0.62V bias.
}
\label{fig2}
\end{figure*}

The data in Figure\,\ref{fig1} show that the mean MTJ resistance at zero field is bias dependent. To probe for faster dynamics, the real time MTJ state is read with an oscilloscope (Methods).  Figure\,\ref{fig2}a shows the signal as a function of time for three different biases. At 0.62\,V, the oscilloscope read time series $V\left( t \right)$ has larger amplitude than the measurement noise shown as 0.0\,V of MTJ bias. Comparatively, at -0.62\,V we observe a signal that is at a similar level as the 0.0\,V bias noise floor. This suggests that the signal increase seen at +0.62\,V originates from MTJ magnetoresistance (MR), and that the dynamics determining MR time series amplitude is asymmetric against MTJ bias voltage.  

To establish the time-scale for randomness in such fluctuating signals, we evaluate the time domain oscilloscope readout's autocorrelation in discrete time-step as
\begin{equation}
R_\mathrm{VV} \left( t_k \right) = \sum_{l=0}^{N-1} V \left( t_l \right) V^*\left( t_{l-k+N-1}\right)
\label{A0}\end{equation}
where $N=\| V \|$ is the length in time-step numbers of time-series voltage signal $V\left( t \right)$, with $t_l = l \Delta t$ representing integer time-steps of the digital scope, with $V\left( t \right)$ set to zero for $t<0$ or $ t>N\Delta t$ for evaluation in Eq.\,\ref{A0}.\footnote{See, for example: https://docs.scipy.org/doc/scipy/reference/ generated/scipy.signal.correlate.html} The width of the resulting $R_\mathrm{VV} \left( t \right)$ near $t=0$ is used as a measure for the time-scale of $V\left( t \right)$'s stochasticity\cite{2019122,2020088}

Figure\,\ref{fig2}b gives an example of such autocorrelation, normalized at its maximum, for a time-series of 10\,$\mu s$ in total length at 50\,ps time-step, at an MTJ bias of +0.62\,V. We extract the width $\Delta \tau_{\mathrm{ac}}$ of the autocorrelation numerically, based on 10\% to 90\% of integrated intensity, corresponding to the grey region in the time-axis in Fig.\,\ref{fig2}b. Multiple bias voltages are used for such time traces and autocorrelation curves. The resulting autocorrelation width's bias dependence is shown in Fig.\,\ref{fig2}c.  For low and negative bias, the autocorrelation is dominated by the noise floor of our measurement system, as evidenced by its apparent independence on bias-voltage across the MTJ. For $V_\mathrm{bias}\gtrsim +0.3$ V, on the other hand,  a systematic change is seen in the autocorrelation width, accompanied by the  increase in signal amplitude of $V\left( t\right)$  as in Fig.\,\ref{fig2}a.

To clarify the bias-voltage regions where the autocorrelation time could be reasonably assigned to MTJ signal, Figure\,\ref{fig2}c shows as a function of MTJ bias voltage both the width of the autocorrelation function on the left-y axis, and the scope-recorded $V\left( t \right)$ swing amplitude on the right y-axis. The signal swing is defined by $V\left( t\right)$'s 2\% and 98\% boundary from its cumulative distribution at a given MTJ bias voltage. Note that $V\left( t \right)$'s mean is artificially leveled by the bias-T isolation and by the low-frequency cutoff of the RF amplifier.  For biases below $\sim +0.2$\,V and negative, we observe little change in $V\left( t \right)$ amplitude, where the time series' autocorrelation width appears dominated by the intrinsic noise characteristics of the measurement setup. For $V_\mathrm{bias}\gtrsim $+0.3\,V a clear increase is seen in the signal amplitude, together with an initial increase of the autocorrelation width. As bias is increased further, the correlation time starts to decrease again. 

The signal amplitude increase for $V_\mathrm{bias}\gtrsim+0.3$ V suggests the measured time series $V\left( t \right)$ is becoming more dominated by MTJ resistance fluctuation above system noise. Therefore, the resulting autocorrelation width change in Fig.\,\ref{fig2}c has its origin in the dynamics of MTJ magnetic moment fluctuation under spin-torque bias. We will compare these observed $\Delta \tau_{\mathrm{ac}}$ bias dependence with modeling analysis later in this study and in supplemental note 1.  

To visualize the nature of stochasticity in observed time series $V\left( t \right)$, lag plots\footnote{ see https://www.itl.nist.gov/div898/handbook/eda/section3/ lagplot.htm} of representative 500\,ns time traces are shown in Fig.\ref{fig2} at different $V_\mathrm{bias}$ points. The autocorrelation width, such as in Fig.\,\ref{fig2}c, suggests that after $\sim$5\,ns the state of the MTJ should be independent of its prior. If so, the resulting lag plot of the delayed MTJ state $V\left( t-5\,\mathrm{ns}\right)$ vs $V\left( t \right)$ would resemble an uncorrelated MTJ resistance fluctuation that visits the full space of possible states. This is indeed what one sees with our time series' lag plots at high positive biases in Fig.\ref{fig2}f-g, where a square-shaped cluster or one bounded by signal-voltage-high is seen, corresponding to the full fluctuation of the MTJ's signal, bounded by its magnetoresistances in parallel (P) and antiparallel (AP) states between the free-layer (FL) and reference-layer (RL).

At lower applied bias, a smaller amplitude, more rounded distribution of values is seen. This represents a cross-over into where measurement noise in the setup starts to dominate. At large negative bias, a slightly larger region of space is occupied, with the distribution relatively circular around zero. This is consistent with the MTJ's anti-parallel state being stabilized further by STT, and the fluctuation amplitude and probability out of AP state being suppressed -- thus $V\left( t\right)$ remains limited by measurement noise primarily.  Indeed as shown by our R-H data in Fig.\,\ref{fig1}, these junctions under study have an excess of AP-preferred coupling from the partially SAF-compensated RL. 

Taking these factors together, we conclude that an MTJ magnetoresistance fluctuation related autocorrelation time of  $\sim$ 2 to 5 ns (depending on bias) has been observed from such MTJ devices as shown here, as expected from earlier model calculations.\cite{2019122,2020088}.

\begin{figure*}[tbp]
\includegraphics[width=\textwidth]{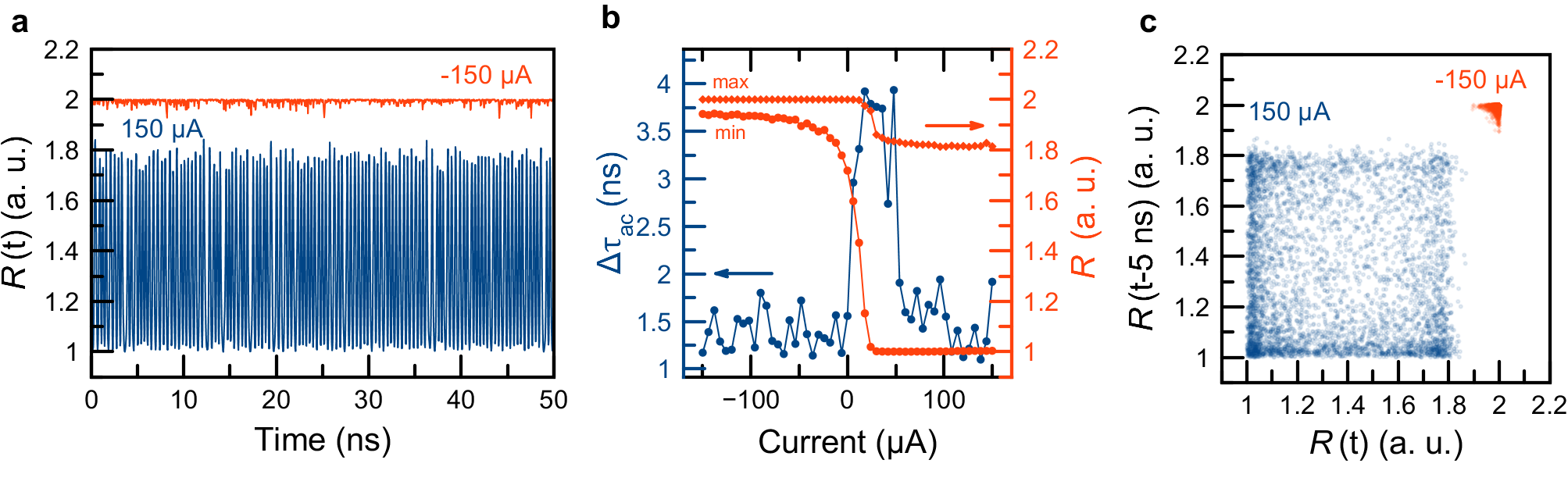}
\caption{ 
(\textbf{a}) Macrospin simulation of the junction state at different spin torque drives. (\textbf{b}) The autocorrelation width and signal amplitudes as a function of bias. (\textbf{c}) Lag plots of the simulated time traces with a 5\,ns delay. 
}
\label{fig3}
\end{figure*} 

Such observed behavior shown in Fig.\,\ref{fig2} is consistent with expected dynamics of an easy-plane superparamagnetic MTJ under thermal agitation. This can be illustrated by comparing these measured data with numerical simulation using a simple, three-moment coupled Landau-Lifshitz-Gilbert equation-set incorporating Langiven fields for finite-temperature fluctuations\cite{2018130}. The FL, top and bottom RL, shown in Fig.\ref{fig1}a, are represented by macrospin moments $m_\mathrm{1,2,3}$ respectively. The LLG equation set includes interlayer coupling. Both dipolar and exchange interactions are included with a simplified effective exchange form. Also included is the Slonczewski-type of antidamping spin torque terms between $m_\mathrm{1,2}$.  Same as in measurements, we define the positive direction of spin-current (i.e. bias current) to be the direction that stabilizes the FL into P-state with respect to the RL. To properly reproduce the observed $R\left(H\right)$ hysteresis loop (see Supplementary note 1) as in Fig.\ref{fig1}b-d, a small (10 to 100 Oe) in-plane uniaxial anisotropy field for $m_\mathrm{1-3}$ is also included. Details of the model's parameter selection and consequences are discussed in Supplementary Note 1.

Figure\,\ref{fig3}a shows the simulated junction tunnel magnetoresistance time series $R\left( t \right)$ over a 50\,ns interval. For negative STT-bias as well as zero (not shown), the junction remains mostly in the AP state with minor thermal fluctuations, consistent with the experimental situation where an imbalance of the SAF reference layer stabilizes the AP state. A positive bias on the other hand, enhances the fluctuation out of AP state towards P. Applying the same numerical procedure as with experiment to these simulated time series, the autocorrelation of the simulated traces is obtained. Figure\,\ref{fig3}b shows the autocorrelation time (left y-axis) together with time series amplitude (right y-axis) as a function of STT bias current. These simulated autocorrelation width agrees with the experimentally observed width of $\sim$ 2 to 5 ns in our MTJ. As observed experimentally, at high, positive direction STT-bias, a decrease of autocorrelation time appears, as the MTJ fluctuates more out of the AP state towards P.  

The autocorrelation time of $\sim$ 2 to 5 ns thus estimated does indeed reflect the time-scale for achieving stochasticity. As done with experimental data, the simulated time series are also presented in lag plots form in Figure\,\ref{fig3}c. For -150\,$\mu A$ of equivalent spin-current, the junction remains mostly in the AP state (red points), while the states it fluctuate out of appears uncorrelated. Simulation of positive voltages at +150\,$\mu A$ shows a square cluster, similar to experimental results, indicating that the simulated junction MR excursion covers nearly the full range of AP and P states with no apparent correlation. 
 
We note that these devices selected for testing have partial interface-mediated perpendicular anisotropy\cite{2010030,2010058,2011014}, which significantly reduces the easy-plane anisotropy field of the FL from that of its full demagnetization value of $4\pi M_\mathrm{s}$. Indeed, to establish the simulation time scale consistent with our observations, we need to directly measure the easy-plane anisotropy field on the same device. This is done using spin-transfer-torque ferromagnetic resonance (STT-FMR)\cite{2005115,2008009B}. For the device shown in Figs.\,\ref{fig1},\ref{fig2}, the directly measured FL easy-plane anisotropy is 1.7\,kOe, and not the full $4\pi M_{s1}$ value of likely in excess of 10\,kOe. This choice of materials parameters is fortuitous, as it allows for better signal to noise ratio in our measurement setup with limited bandwidth and with impedance mismatch. Increases in fluctuation speed should be obtainable by decreasing perpendicular anisotropy.

As a further test for the stochastic nature of our MTJ signal as a true random number generator, we convert the MTJ's analog signal time series into a binary digital series, and test their stochasticity by applying the National Institute of Standards and Technology Statistical Test Suite (NIST STS) \cite{bassham_iii_sp_2010}(See Supplementary Note 2). The  analog time series signal measured at 0.62\,V (as illustrated in Fig.\,\ref{fig2}a) is transformed to a random bitstream of 0s and 1s (the so-called Bernoulli sequence) by thresholding the signal at its analog median. This procedure ensures proper probability distribution of the resulting Bernoulli series at 50\% probability for the NIST STS tests, which is important to achieve high test performance. For this test, the stochastic MTJ signal is measured for a total of $8 \times 10^6$ sampling points at different sampling frequencies. XOR-combined Bernoulli series are also generated to whiten the distribution for reference, with this method being a common approach to improve the quality of the random bitstream \cite{2017150}. In this process, the measured sequences are divided into 8 sub-sequences of $10^6$ sampling points each. The NIST STS test is run for the directly converted Bernoulli series and its XOR-derived series for different sampling times $t_{\rm samp}$ for a total of $10^6$ sampling points each. For the XOR$n$ processed series, the number $n$ indicates how many independent bitstream are XOR combined. The 188 tests with 15 different tests types that are part of the NIST STS are run. The percentage of tests passed for the different configurations is shown in Fig.\,\ref{fig4}. More information about what test types are passed for each configuration is described in Supplementary Note 2.
\begin{figure}[tbp]
\includegraphics[width=0.5\textwidth]{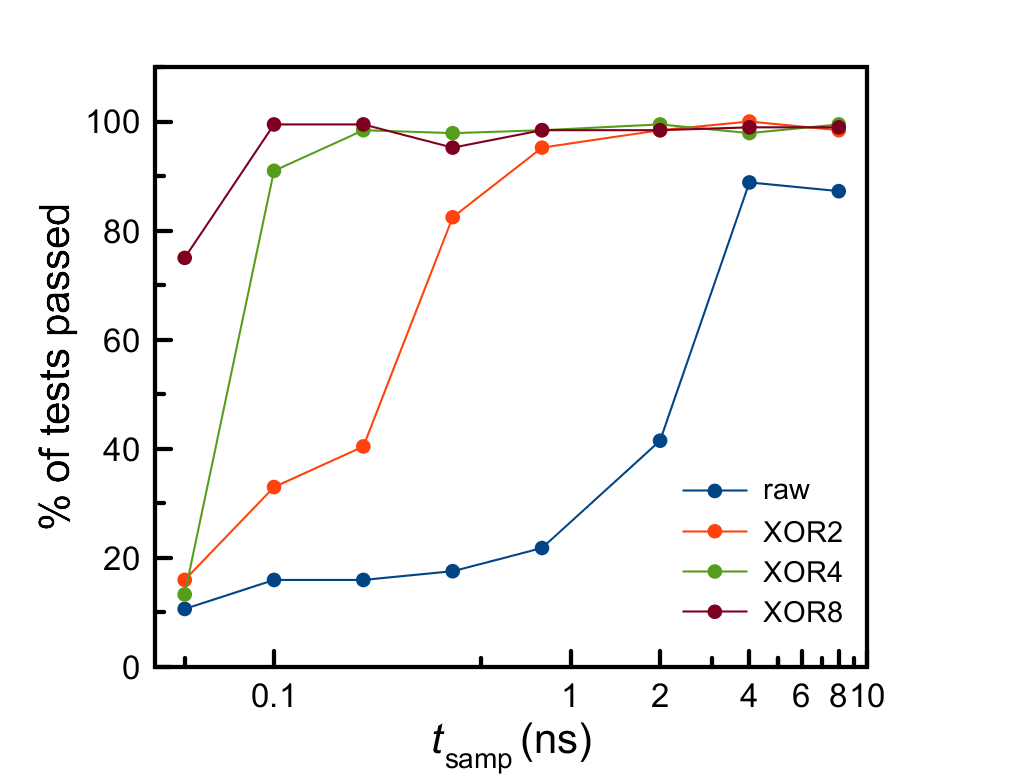}
\caption{Percentage of tests passed of the NIST STS for the raw Bernoulli series and XOR2, XOR4, XOR8 for different sampling times $t_{\rm samp}$. Raw Bernoulli series is obtained from the MTJ fluctuation signal at +0.62\,V bias, as illustrated in Fig.\ref{fig2}a.}
\label{fig4}
\end{figure}

It can be clearly seen that larger sampling times, as well as the whitening of the signal by multiple XOR combinations, improve the quality of the randomness of the stochastic bitstream. For applications like true random number generators for probabilistic computing, an important criteria is the limit of how fast we can sample and still obtain a completely uncorrelated random bit. The test performance for the raw digital series reproduces a time scale of 2 to 5\,ns, similar to the auto correlation length of the analog signal discussed above. For XOR2 (from two series with one XOR-combine operation), the quality of the stochastic signal is close to 100\% for sampling times of $t_{\rm samp} \geq 1$ ns. The test performance is further improved for  XOR4 and XOR8. This shows that the stochastic MTJ can generate good quality random numbers on a nanosecond rate. 

We have to note that in the context of the NIST STS, we have analyzed the stochastic MTJ signal for the case that the Bernoulli sequence is balanced with equal numbers of 0s and 1s. For some applications in the context of Ising and probabilistic computing, tunability of the random sequence's mean-value is important. In our experiment, tunability can for example be achieved by moving the threshold value for binarization away from the median or by varying the bias voltage $V_\mathrm{bias}$ at a fixed threshold value. In hardware, compact tunable random number generator designs or p-bits incorporating stochastic MTJs have been presented by references \cite{camsari_stochastic_2017,camsari_implementing_2017,2019122}. Further analysis of a biased Bernoulli sequence is beyond  the scope of this paper.

In conclusion, we provide experimental evidence that for easy-plane MTJs with very small in-plane anisotropy, the thermal fluctuation related magnetoresistance time series indeed show stochasticity with an autocorrelation time below 5\,ns,  dictated by the easy-plane anisotropy field.  Our experimental results are consistent with prior expectation from theoretical models, and with our own numerical simulation. Easy-plane superparamagnetic MTJ could therefore be a viable device component for probabilistic computing and related applications.

\section*{Methods}

MTJs are patterned from sputtered films based  of $\parallel$ Bottom contact $\mid$ 20 CoFeB $\mid$ 8 Ru $\mid$ 30 CoFeB $\mid$ ~10 MgO $\mid$ ~ 30 CoFeB $\mid$ Top contact $\parallel$ (numbers are for thickness, in \AA). They are patterned into junction devices using photolithography and Ar ion milling techniques. The completed junction diameter is 60\,nm. 

Electrical measurements are performed at a constant voltage bias provided through the DC side of a bias tee.  MTJ resistance is calculated from the current flowing through a sense resistor in series with the MTJ. The high frequency behavior is measured through the \textit{rf} side of the bias tee. The MTJ signal is amplified by a 26\,dB broadband amplifier (30\,kHz to 12\,GHz) and measured by a digital oscilloscope with a 20\,Gs/s max sampling rate. All measurements reported are performed at ambient temperature of about 20\,C.

\section*{Acknowledgment}

We acknowledge fruitful discussions with our colleagues, in particular, with Eric Edwards on MTJ wafer-level characteristics and Supriyo Datta on p-bit design.  Work was done with the MRAM group at IBM T. J. Watson Research Center in Yorktown Heights, New York. J.K. also acknowledges the support of an IBM Summer Internship program that made this study possible.

\bibliographystyle{naturemag}

\bibliography{main_full}

\clearpage
\widetext

\newcommand{\uvec}[1]{\hat{\mathbf{#1}}}
\newcommand{\plane}[1]{#1}

\renewcommand{\figurename}{Supplementary Figure}
\renewcommand\refname{Supplementary References}
\renewcommand{\tablename}{Supplementary Table}
\def\bibsection{\section*{\refname}} 

\begin{center}

{\Large \textbf{Supplementary Material for \\
Demonstration of nanosecond operation in stochastic magnetic tunnel junctions}}\\
\end{center}
\bigskip
\onecolumngrid
\setcounter{equation}{0}
\setcounter{figure}{0}
\setcounter{table}{0}
\setcounter{page}{1}
\renewcommand{\theequation}{S\arabic{equation}}
\renewcommand{\thefigure}{S\arabic{figure}}

\renewcommand{\figurename}{Supplementary Figure}
\renewcommand\refname{Supplementary References}
\def\bibsection{\section*{\refname}} 

\section*{Supplementary Note 1: Three moment coupled macrospin model definition and magnetoresistance}

These observed behavior can be understood semi-quantitatively within a simple, three-moment coupled Landau-Lifshitz-Gilbert equation-set incorporating Langiven fields for finite-temperature fluctuations\cite{2018130}. The three layers shown in Fig.1(a) are reprsented by three macrospins $m_\mathrm{1,2,3} = M_\mathrm{s1,2,3} t_\mathrm{1,2,3} a^2$, with $t_\mathrm{1,2,3}$ representing the thicknesses, and $a$ the lateral size of the device pillar.\footnote{The factor $\pi/4$ is taken as one for simplicity without loss of generality.} A Slonczewski-type of ``anti-damping'' spin-torque\cite{98114,2005073} is included between the FL moment $m_1$  and top RL moment $m_2$. The RL $m_2$ is exchange coupled to  the bottom reference layer $m_3$ with energy $E_\mathrm{ex23}$. One assumes $m_\mathrm{1,2,3}$ are collinear in the film normal direction, and are in their easy-plane anisotropy limit, with a small in-plane uniaxial anisotropy for each, also a co-axial fashion in the film plane. The small in-plane anisotropy is necessary to account for imperfections in MTJ pillar shape definition as well as possible crystalline anisotropies formed during the in-field annealing of the film stack.  The three moments are further coupled together via dipolar fields among them. A model thus defined can be written as
\begin{equation}
\left\{
\begin{array}{l}
\dfrac{d\mathbf{n}_{m1}}{dt}=\gamma \mathbf{H}_\mathrm{total1}
   \times \mathbf{n}_{m1} +\alpha _1  \mathbf{n}_{m1} \times \dfrac{d\mathbf{n}_{m1}}{dt} +\left( \dfrac{J_\text{s}}{M_{s1}t_1}\right) \mathbf{n}_{m1} \times \left(\mathbf{n}_{m1}\times \mathbf{n}_{m2} \right) \\
\\
\dfrac{d\mathbf{n}_{m2}}{dt}=\gamma  \mathbf{H}_\mathrm{total2}
   \times \mathbf{n}_{m2} +\alpha _2  \mathbf{n}_{m2} \times \dfrac{d\mathbf{n}_{m2}}{dt} -\left( \dfrac{J_\text{s}}{M_{s2}t_2}\right) \mathbf{n}_{m2} \times \left(\mathbf{n}_{m2}\times \mathbf{n}_{m1} \right) \\
\\
\dfrac{d\mathbf{n}_{m3}}{dt}=\gamma \mathbf{H}_\mathrm{total3}
   \times \mathbf{n}_{m3} +\alpha _3 \mathbf{n}_{m3} \times \dfrac{d\mathbf{n}_{m3}}{dt}\\
\end{array}
\right.\label{E1}
\end{equation}
with 
\begin{equation}
\left\{
\begin{array}{l}
\mathbf{H}_\mathrm{total1}= - H_\mathrm{p1}\left( \mathbf{n}_\mathrm{p1}\cdot \mathbf{n}_\mathrm{m1}\right)\mathbf{n}_\mathrm{p1} +H_\mathrm{k1} \mathbf{n}_\mathrm{k1}+\dfrac{E_\mathrm{ex12}}{M_\mathrm{s1} t_1}\mathbf{n_\mathrm{m2}} +\dfrac{E_\mathrm{ex13}}{M_\mathrm{s1} t_1}\mathbf{n_\mathrm{m3}} +\mathbf{H}_\mathrm{a} +\mathbf{H}_\mathrm{L1} \\
\mathbf{H}_\mathrm{total2} = -H_\mathrm{p2}\left( \mathbf{n}_\mathrm{p2}\cdot \mathbf{n}_\mathrm{m2}\right)\mathbf{n}_\mathrm{p2} +H_\mathrm{k2} \mathbf{n}_\mathrm{k2}+\dfrac{E_\mathrm{ex12}}{M_{s2}t_2} \mathbf{n}_\mathrm{m1}+ \dfrac{E_\mathrm{ex23}}{M_{s2}t_2} \mathbf{n}_\mathrm{m3}+\mathbf{H}_\mathrm{a} + \mathbf{H}_\mathrm{L2}\\
\mathbf{H}_\mathrm{total3} = -H_\mathrm{p3}\left( \mathbf{n}_\mathrm{p3}\cdot \mathbf{n}_\mathrm{m3}\right)\mathbf{n}_\mathrm{p3} +H_\mathrm{k3} \mathbf{n}_\mathrm{k3}+\dfrac{E_\mathrm{ex13}}{M_{s3}t_3} \mathbf{n}_\mathrm{m1}+ \dfrac{E_\mathrm{ex23}}{M_{s3}t_3} \mathbf{n}_\mathrm{m2}+\mathbf{H}_\mathrm{a} + \mathbf{H}_\mathrm{L3}
\end{array}
\right.
\label{E2}
\end{equation}
where (with $i \in 1,2,3$)  $\mathbf{n}_\mathrm{mi}$ are unit vectors for the directions of moment $m_\mathrm{i}$,  $\mathbf{n}_\mathrm{pi}$ for the directions of easy-plane norm of each layers (set collinear to $\mathbf{e}_\mathrm{z}$ the z-direction); $H_\mathrm{pi}$ the easy-plane anisotropy field,
$H_\mathrm{ki}$ and $\mathrm{n}_\mathrm{ki}$ the in-plane uniaxial anisotropy's easy-axes for each (assumed collinear along $\mathbf{e}_\mathrm{x}$), and $E_\mathrm{exij}$ with $i,j \in 1,2,3$ the exchange-energy among the respective moments. Here $E_\mathrm{exij}$ includes both interlayer exchange coupling and an approximate dipolar coupling. $\mathbf{H}_\mathrm{L}$s are Langevin fields for thermal fluctuation as defined by
\begin{equation}
\left\{
\begin{array}{l}
\left< H_\mathrm{Li(u,v,w)} \right> =0 \\ \\
\left< H_\mathrm{Liu}\left(t-\tau\right) H_\mathrm{Liv}\left(t\right) \right>= 2 D_\mathrm{hi} \delta_\mathrm{u,v}\delta\left(\tau \right)\\
\end{array}\right.
\label{E3}\end{equation}
where $u,v,w \in 1,2,3$ representing  $\mathbf{e}_\mathrm{x,y,z}$ the three Cartesian components, $i \in 1,2,3$ for the three moments $m_\mathrm{i}$, and $D_\mathrm{hi}$ the corresponding Fokker-Planck diffusion constants from fluctuation-dissipation constraint as\cite{2018130}
\begin{equation}
D_\mathrm{hi} =
\dfrac{ \alpha_i k_B T}{\left( 1 + \alpha_i^2\right) \gamma M_\mathrm{si} t_i a^2}
\label{E4}\end{equation}
$\alpha_\mathrm{i}$ are the LLG-damping factor for the moments; $J_\mathrm{s}$ is the spin current flowing through the layer in units of magnetic moment, generating spin-torque across the interface between $m_1$ and $m_2$. For simplicity, we assume equal charge to spin conversion efficiency for both layers.    $\gamma \approx 2 \mu_B/\hbar$ is the magnitude of the gyro-magnetic ratio, and $\mathbf{H}_a$ is the applied field, in $\mathbf{e}_x$ direction, collinear with the in-plane uniaxial anisotropy axis. A representative set of parameters used in the numerical simulation is shown in Supplementary Table\,\ref{sim1_table}. 

\begin{table*}[]
\begin{tabular}{ccccccccc}
   \hline
   \hline
   Moment & $M_{\mathrm{s}}$ ($emu\,cm^{-3}$)  & $H_p $ (Oe) & $H_{k}$ (Oe) & $\alpha$ & t (nm)  & $ H_\mathrm{ex}$\footnote{Here one writes $\left(H_\mathrm{ex}\right)_\mathrm{ij}$ so that $E_\mathrm{ij}=E_\mathrm{ji}=\left(H_\mathrm{ex}\right)_\mathrm{ij} m_\mathrm{j}=\left(H_\mathrm{ex}\right)_\mathrm{ji} m_\mathrm{i}$, with $i,j  \in \left[1,2,3\right]$ for $m_\mathrm{1,2,3}$ respectively.} & $H_\mathrm{k}$ (Oe)\\
   \hline
$m_1$ & 800 & 1700   & 50 &0.005  & 1.25 & $\left(-900\right)_{21}$, $\left(-725\right)_{31}$ &50 \\ 
$m_2$ & 1400 &$0.5\times 4\pi M_\mathrm{s2}$ &100 & 0.03 & 3 & $\left(-1000\right)_{32}$  & 100 \\
$m_3$ & 1435 & $0.75\times 4\pi M_\mathrm{s3}$& 100 & 0.03   &  3 & & 100  \\
\hline
\hline
\end{tabular}
\caption{Material parameters used in the three macrospin-moment model .}
\label{sim1_table}
\end{table*}

A representative set of simulation results is shown in Supplemental Fig.\ref{figfs1} for an MTJ's magnetoresistance dependence on applied field. Only data from up-sweep of magnetic field is shown, and with the magnetic field applied in the direction of the weak in-plane anisotropy. These traces reproduce the main features observed in our experimental device in Fig.1, and is in semi-quantitative agreement with Fig.1. Here we only simulate the magnetodynamics with stochastic LLG, the magnetoresistance is taken from the inner product $\mathbf{n}_\mathrm{m1}\cdot \mathbf{n}_\mathrm{m2}$ in the form of $R_\mathrm{MTJ}\left[H_a\right]=\left<4/\left(3+  \mathbf{n}_\mathrm{m1}\cdot \mathbf{n}_\mathrm{m2}\right)\right>$, where the $\left< ...  \right>$ is averaging taken over the 200 ns simulated time series length (excluding the initial 10 ns to avoid any residual initial state transient). The values in $R\left[H_a\right]$ are chosen so that $R_\mathrm{min}=1$ and $R_\mathrm{max}=2$, representing the full swing of a tunnel magnetoresistance of 100\%.
\begin{figure*}[tbp]
\includegraphics[width=6in]{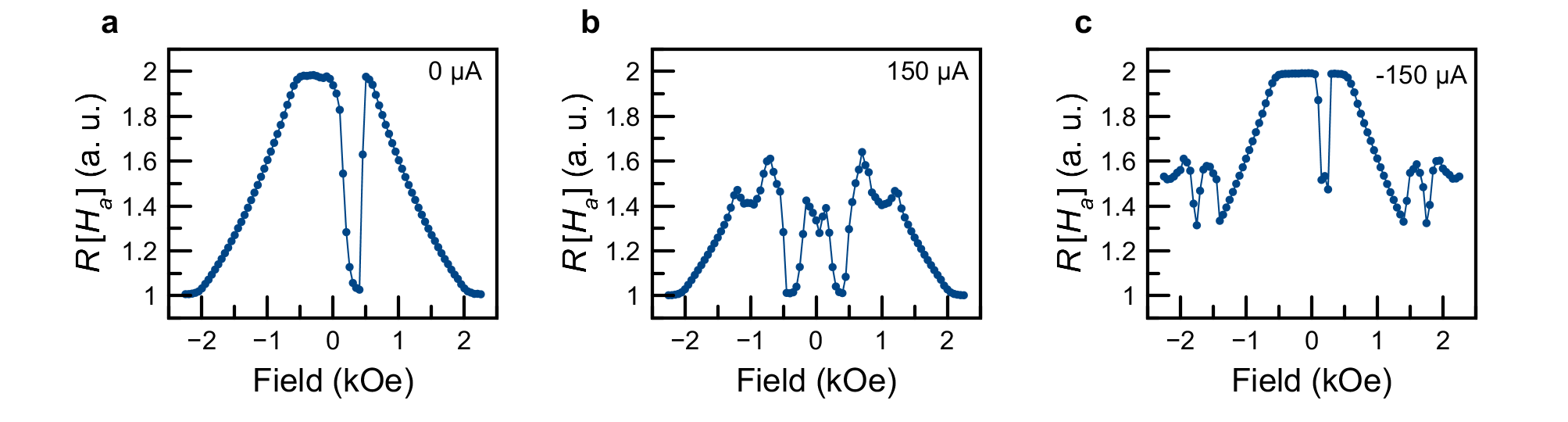}
\caption{ 
Simulation generated magnetoresistance vs applied field. The field is applied in the direction of the weak in-plane easy-axis. Only data for field-increasing sweep are shown. The spin-current applied (in charge current unit) is shown in labels.}
\label{figfs1}
\end{figure*} 

The corresponding time domain behavior is shown in the main text as Fig.3. 

\section*{Supplementary Note 2: Statistical test of the stochastic signal generated by the MTJ}
In this section, additional information about the results of the NIST STS described in the main manuscript are given. Table \ref{table: NIST raw} shows the passed tests for the raw signal with $10^6$ sampling points for different sampling times $t_{\rm samp}$.  Table \ref{table: NIST XOR8} contains the same information for the case where 8 independent bitstreams are XOR-combined. A total of 188 tests are run where 148 tests belong to the test type 'Non-overlapping Template Matching'. This test type searches the Bernoulli sequences for 148 predefined templates of bit length 9 and tests if these templates occur too often or too little. This test type dominates the overall numbers of tests. The remaining 40 tests are divided between the 14 remaining tests where some tests types consist of more than one test. In Fig. 4 of the main manuscript, the number of passed tests is divided by the number of all tests to obtain the percentage of tests passed.

As shown in Table \ref{table: NIST raw} for the raw signal, only the Binary-Matrix-Rank test is passed for all sampling frequencies. However, most tests are passed for times $\geq2$ ns. For the XOR8 signal shown in Table \ref{table: NIST XOR8} all test are (nearly) passed for sampling times of $t_{\rm samp} \geq 0.1$ ns. This shows that the stochastic MTJs can generate random numbers of cryptographic quality on a nanosecond rate.

In the context of Ising and probabilistic computing where compact hardware implementations of entropy sources are desired, the performance of a raw signal belonging to a single stochastic MTJ is important since additional overhead in form of XOR gates or similar should ideally be avoided. While our results show that the raw signal does not achieve perfect performance in terms of cryptographic quality according to the NIST STS, the requirements on the quality of randomness for Ising and probabilistic computing differ from the NIST STS. The observed fluctuations characteristics from the stocastic MTJ should hence be sufficient to achieve good system level performance on tasks like optimization. This shows that the stochastic MTJ signal without any post-processing should be sufficient to be utilized for applications in Ising or probabilistic computing.

\begin{table}
\begin{center}
 \begin{tabular}{ c |  c  c  c  c  c  c c c } \hline
 \hline
Tests \ \textbackslash \ $t_{\rm samp}$ & $0.05$ ns & $0.1$ ns & $0.2$ ns & $0.4$ ns & $0.8$ ns & $2.0$ ns & $4.0$ ns & $8.0$ ns  \\  
 \hline
Frequency  & \cellcolor{red} 0/1 & \cellcolor{red} 0/1 & \cellcolor{red} 0/1 & \cellcolor{red} 0/1 & \cellcolor{red} 0/1 & \cellcolor{red} 0/1 &  \cellcolor{red} 0/1 & \cellcolor{red} 0/1  \\  
Block Frequency   & \cellcolor{red} 0/1 & \cellcolor{red} 0/1 & \cellcolor{red} 0/1 & \cellcolor{red} 0/1 & \cellcolor{red} 0/1 & \cellcolor{red} 0/1 &  \cellcolor{red} 0/1 & \cellcolor{red} 0/1  \\  
Runs   & \cellcolor{red} 0/1 & \cellcolor{red} 0/1 & \cellcolor{red} 0/1 & \cellcolor{red} 0/1 & \cellcolor{red} 0/1 & \cellcolor{red} 0/1 &  \cellcolor{red} 0/1 & \cellcolor{red} 0/1  \\   
Longest-Runs-of-Ones  & \cellcolor{red} 0/1 & \cellcolor{red} 0/1 & \cellcolor{red} 0/1 & \cellcolor{red} 0/1 & \cellcolor{green} 1/1 & \cellcolor{red} 0/1 &  \cellcolor{green} 1/1 & \cellcolor{green} 1/1  \\  
 Binary-Matrix-Rank   & \cellcolor{green} 1/1 & \cellcolor{green} 1/1 & \cellcolor{green} 1/1& \cellcolor{green} 1/1 & \cellcolor{green} 1/1 & \cellcolor{green} 1/1 & \cellcolor{green} 1/1 &  \cellcolor{green} 1/1  \\  
DFFT   & \cellcolor{red} 0/1 & \cellcolor{red} 0/1 & \cellcolor{red} 0/1&  \cellcolor{red} 0/1 & \cellcolor{red} 0/1 & \cellcolor{red} 0/1  & \cellcolor{green} 1/1  & \cellcolor{green} 1/1  \\  
Non-overlapping Template Matching   & \cellcolor{red} 0/148 & \cellcolor{yellow} 2/148 & \cellcolor{yellow} 2/148 & \cellcolor{yellow} 5/148 & \cellcolor{yellow} 19/148 & \cellcolor{yellow} 50/148 &  \cellcolor{yellow} 134/148 &  \cellcolor{yellow} 135/148  \\  
Overlapping Template Matching    & \cellcolor{red} 0/1 & \cellcolor{red} 0/1 & \cellcolor{red} 0/1&  \cellcolor{red} 0/1 & \cellcolor{red} 0/1 & \cellcolor{red} 0/1  & \cellcolor{green} 1/1  & \cellcolor{green} 1/1  \\    
Maurer's Universal Statistical Test    & \cellcolor{red} 0/1 & \cellcolor{red} 0/1 & \cellcolor{red} 0/1&  \cellcolor{red} 0/1 & \cellcolor{red} 0/1 & \cellcolor{red} 0/1  & \cellcolor{red} 0/1  & \cellcolor{green} 1/1  \\  
Linear Complexity    & \cellcolor{green} 1/1 & \cellcolor{green} 1/1 & \cellcolor{green} 1/1& \cellcolor{green} 1/1 & \cellcolor{green} 1/1 & \cellcolor{green} 1/1 & \cellcolor{green} 1/1 &  \cellcolor{green} 1/1  \\  
Serial   &\cellcolor{red} 0/2 & \cellcolor{red} 0/2 & \cellcolor{red} 0/2 & \cellcolor{red} 0/2 & \cellcolor{yellow} 1/2 & \cellcolor{yellow} 1/2 &  \cellcolor{green} 2/2 &  \cellcolor{yellow} 1/2  \\  
Approximate Entropy   & \cellcolor{red} 0/1 & \cellcolor{red} 0/1 & \cellcolor{red} 0/1 & \cellcolor{red} 0/1 & \cellcolor{red} 0/1 & \cellcolor{red} 0/1 &  \cellcolor{red} 0/1 & \cellcolor{red} 0/1  \\  
Cumulative sum  & \cellcolor{red} 0/2 & \cellcolor{red} 0/2 & \cellcolor{red} 0/2 & \cellcolor{red} 0/2 & \cellcolor{red} 0/2 & \cellcolor{red} 0/2 &  \cellcolor{red} 0/2 & \cellcolor{red} 0/2  \\  
Random Excursion & \cellcolor{red} 0/8 & \cellcolor{green} 8/8 & \cellcolor{green} 8/8 & \cellcolor{green} 8/8 & \cellcolor{yellow} 7/8 & \cellcolor{yellow} 7/8 &  \cellcolor{green} 8/8 & \cellcolor{green} 8/8  \\  
Random Excursion Variant & \cellcolor{green} 18/18 & \cellcolor{green} 18/18 & \cellcolor{green} 18/18 & \cellcolor{green} 18/18 & \cellcolor{yellow} 11/18 & \cellcolor{green} 18/18 &  \cellcolor{green} 18/18 & \cellcolor{yellow} 14/18  \\  
 \hline \hline

\end{tabular}
\end{center}
\caption{NIST STS results for the raw data for different sampling times $t_{\rm samp}$.}
\label{table: NIST raw}
\end{table}

\begin{table}
\begin{center}
 \begin{tabular}{ c |  c  c  c  c  c  c c c } \hline
 \hline
Tests \ \textbackslash \ $t_{\rm samp}$ & $0.05$ ns & $0.1$ ns & $0.2$ ns & $0.4$ ns & $0.8$ ns & $2.0$ ns & $4.0$ ns & $8.0$ ns  \\  
 \hline
Frequency  & \cellcolor{green} 1/1 & \cellcolor{green} 1/1 & \cellcolor{green} 1/1& \cellcolor{green} 1/1 & \cellcolor{green} 1/1 & \cellcolor{green} 1/1 & \cellcolor{green} 1/1 &  \cellcolor{green} 1/1  \\  
Block Frequency    & \cellcolor{red} 0/1 & \cellcolor{green} 1/1 & \cellcolor{green} 1/1& \cellcolor{green} 1/1 & \cellcolor{green} 1/1 & \cellcolor{green} 1/1 & \cellcolor{green} 1/1 &  \cellcolor{green} 1/1  \\  
Runs   & \cellcolor{red} 0/1 & \cellcolor{green} 1/1 & \cellcolor{green} 1/1& \cellcolor{green} 1/1 & \cellcolor{green} 1/1 & \cellcolor{green} 1/1 & \cellcolor{green} 1/1 &  \cellcolor{green} 1/1  \\  
Longest-Runs-of-Ones  & \cellcolor{red} 0/1 & \cellcolor{green} 1/1 & \cellcolor{green} 1/1& \cellcolor{green} 1/1 & \cellcolor{green} 1/1 & \cellcolor{green} 1/1 & \cellcolor{green} 1/1 &  \cellcolor{green} 1/1  \\  
 Binary-Matrix-Rank   & \cellcolor{green} 1/1 & \cellcolor{green} 1/1 & \cellcolor{green} 1/1& \cellcolor{green} 1/1 & \cellcolor{green} 1/1 & \cellcolor{green} 1/1 & \cellcolor{green} 1/1 &  \cellcolor{green} 1/1  \\  
DFFT   & \cellcolor{green} 1/1 & \cellcolor{green} 1/1 & \cellcolor{green} 1/1& \cellcolor{green} 1/1 & \cellcolor{green} 1/1 & \cellcolor{green} 1/1 & \cellcolor{green} 1/1 &  \cellcolor{green} 1/1  \\  
Non-overlapping Template Matching   & \cellcolor{yellow} 107/148 & \cellcolor{yellow} 147/148 & \cellcolor{yellow} 147/148 & \cellcolor{yellow} 145/148 & \cellcolor{yellow} 145/148 & \cellcolor{yellow} 145/148 &  \cellcolor{yellow} 146/148 &  \cellcolor{yellow} 146/148  \\  
Overlapping Template Matching    & \cellcolor{red} 0/1 & \cellcolor{green} 1/1 & \cellcolor{green} 1/1& \cellcolor{green} 1/1 & \cellcolor{green} 1/1 & \cellcolor{green} 1/1 & \cellcolor{green} 1/1 &  \cellcolor{green} 1/1  \\    
Maurer's Universal Statistical Test    & \cellcolor{green} 1/1 & \cellcolor{green} 1/1 & \cellcolor{green} 1/1& \cellcolor{green} 1/1 & \cellcolor{green} 1/1 & \cellcolor{green} 1/1 & \cellcolor{green} 1/1 &  \cellcolor{green} 1/1  \\  
Linear Complexity     & \cellcolor{green} 1/1 & \cellcolor{green} 1/1 & \cellcolor{green} 1/1& \cellcolor{green} 1/1 & \cellcolor{green} 1/1 & \cellcolor{green} 1/1 & \cellcolor{green} 1/1 &  \cellcolor{green} 1/1  \\  
Serial   &\cellcolor{yellow} 1/2 & \cellcolor{green} 2/2 & \cellcolor{green} 2/2 & \cellcolor{green} 2/2 & \cellcolor{green} 2/2 & \cellcolor{green} 2/2 &  \cellcolor{green} 2/2 &  \cellcolor{green} 2/2 \\  
Approximate Entropy   & \cellcolor{red} 0/1 & \cellcolor{green} 1/1 & \cellcolor{green} 1/1& \cellcolor{green} 1/1 & \cellcolor{green} 1/1 & \cellcolor{green} 1/1 & \cellcolor{green} 1/1 &  \cellcolor{green} 1/1  \\  
Cumulative sum  &\cellcolor{green} 2/2 & \cellcolor{green} 2/2 & \cellcolor{green} 2/2 & \cellcolor{green} 2/2 & \cellcolor{green} 2/2 & \cellcolor{green} 2/2 &  \cellcolor{green} 2/2 &  \cellcolor{green} 2/2 \\  
Random Excursion & \cellcolor{green} 8/8 & \cellcolor{green} 8/8 & \cellcolor{green} 8/8 & \cellcolor{yellow} 6/8 & \cellcolor{green} 8/8 & \cellcolor{green} 8/8 &  \cellcolor{green} 8/8 & \cellcolor{green} 8/8  \\  
Random Excursion Variant & \cellcolor{green} 18/18 & \cellcolor{green} 18/18 & \cellcolor{green} 18/18 & \cellcolor{yellow} 14/18 & \cellcolor{green} 18/18 & \cellcolor{green} 18/18 &  \cellcolor{green} 18/18 & \cellcolor{green} 18/18  \\  
 \hline \hline

\end{tabular}
\end{center}
\caption{NIST STS results for XOR8 for different sampling times $t_{\rm samp}$.}
\label{table: NIST XOR8}

\end{table}

%


\end{document}